\documentclass[twocolumn]{aastex63}

\graphicspath{{./}{figs/}}

\usepackage{amsmath}
\usepackage{appendix}

\usepackage{needspace}

\newif\ifref
\reftrue
\reffalse
\definecolor{darkred}{rgb}{0.75, 0, 0}
\newcommand{\mb}[1]{\ifref\textcolor{darkred}{#1}\else #1\fi}

\newif\ifreff
\refftrue
\refffalse
\definecolor{darkred}{rgb}{0.75, 0, 0}
\newcommand{\mbb}[1]{\ifreff\textcolor{darkred}{#1}\else #1\fi}

\newif\ifrefff
\reffftrue
\reffffalse
\definecolor{darkred}{rgb}{0.75, 0, 0}
\newcommand{\mbbb}[1]{\ifrefff\textcolor{darkred}{#1}\else #1\fi}

\accepted{September 12, 2019}

\submitjournal{Astrophysical Journal Letters}

\shorttitle{Asteroseismic Constraints on $\dot G$}
\shortauthors{Bellinger \& Christensen-Dalsgaard}

\begin{document}
\title{\normalfont Asteroseismic constraints on the cosmic-time variation of \\the gravitational constant from an ancient main-sequence star}

\correspondingauthor{E.\ P.\ Bellinger}
\email{bellinger@phys.au.dk}

\author[0000-0003-4456-4863]{Earl Patrick Bellinger} 
\altaffiliation{SAC Postdoctoral Fellow} 
\affil{Stellar Astrophysics Centre\\ 
Department of Physics and Astronomy\\ 
Aarhus University\\ 
Denmark} 

\author[0000-0001-5137-0966]{J{\o}rgen Christensen-Dalsgaard} 
\affil{Stellar Astrophysics Centre\\ 
Department of Physics and Astronomy\\ 
Aarhus University\\ 
Denmark}

\begin{abstract} 
    We investigate the variation of the gravitational constant $G$ over the history of the Universe by modeling the effects on the evolution and asteroseismology of the low-mass star KIC~7970740, which is one of the oldest ($\sim 11$~Gyr) and best-observed solar-like oscillators in the Galaxy. 
    From these data we find ${\dot{G}/G = \mbbb{(1.2 \pm 2.6)} \times 10^{-12}~\text{yr}^{-1}}$, that is, no evidence for any variation in $G$. 
    We also find a Bayesian asteroseismic estimate of the age of the Universe as well as astrophysical S-factors for five nuclear reactions obtained through a 12-dimensional stellar evolution Markov chain Monte Carlo simulation. 
\end{abstract}


\section{Introduction} \label{sec:intro} 
Is the gravitational constant actually constant? 
Interest in this question goes back at least to the time of \citet{1937Natur.139..323D}. 
On the one hand, Einstein's theory of general relativity says \emph{yes}: according to the equivalence principle, the outcome of any local experiment in a freely falling laboratory is independent of its position in spacetime. 
Hence, $G$ is the same everywhere for all time. 
String theory and other theories of modified gravity, on the other hand, say \emph{no}: the gravitational `constant' is rather a derived parameter which can vary over cosmic time (see, e.g., \citealt{2003RvMP...75..403U, Uzan2011} and \citealt{2011PThPh.126..993C} for reviews). 

The constancy of $G$ is an empirical question which can be investigated through astrophysical experimentation. 
The strongest constraints to date come from the dynamics of the solar system. 
The Lunar Ranging Experiment \citep{1962Natur.194.1267S, 2013RPPh...76g6901M} gives $\dot{G}/G = {(7.1\pm7.6) \times 10^{-14}}\;\text{yr}^{-1}$ over the past few decades \citep{0264-9381-35-3-035015}. 
Similarly, the \textsc{Messenger} probe \citep{2018NatCo...9..289G} has used the ephemeris of Mercury to find ${|\dot{G}|/G < 4 \times 10^{-14}}\;\text{yr}^{-1}$ over the past seven years. 
Other local (in both time and space) constraints have been derived from other planetary motions \citep{PhysRevLett.51.1609}, exoplanetary motions \citep{2016PASJ...68L...5M}, and pulsar binaries \citep{1991ApJ...366..501D, 2019MNRAS.482.3249Z}, among others. 

Though these experiments are consistent with a constant $G$, they do not probe $\dot G$ over cosmic time, where presumably any major variations to $G$ would have transpired. 
Experiments which do probe cosmic time, albeit in a model-dependent fashion, include measurements from helioseismology \citep{1998ApJ...498..871G}, white dwarfs \citep{2011JCAP...05..021G, 2013JCAP...06..032C}, and globular clusters \citep{1996A&A...312..345D}. 
More distant constraints have been derived from big bang nucleosynthesis \citep{1990PhLB..248..146A} and anisotropy in the cosmic microwave background \citep{2004PhRvD..69h3512N}. 
These experiments are also consistent with a constant $G$, albeit with greater uncertainty ($|\dot G/G| \lessapprox 10^{-12}\;\text{yr}^{-1}$). 

In this Letter, we contribute a new experiment to test the cosmic-time variation of $G$ using asteroseismology. 
Thanks to four years of observations from the \emph{Kepler} mission \citep{2010Sci...327..977B}, there are now extraordinarily accurate measurements of stellar oscillations from solar-like stars in the Galaxy. 
For a typical well-observed solar-type star, dozens of oscillation mode frequencies can be resolved. 
As the properties of the oscillations depend on the properties of the star, asteroseismic data can be used to constrain stellar global parameters. 
By furthermore assuming that the theory of stellar evolution is approximately correct, constraints can be placed on the age and evolutionary history of the star by fitting models to the data. 

Here we study a rich spectrum of acoustic oscillation mode frequencies measured from a low-mass solar-like star on the main sequence, KIC~7970740, and determine whether the observations of this star are consistent with a constant gravitational constant. 
The use of a low-mass star such as this one is ideal because it avoids the theoretical uncertainties associated with higher mass stars, such as element diffusion and convective core overshoot. 

\begin{figure}
    \centering
    \includegraphics[width=\linewidth]{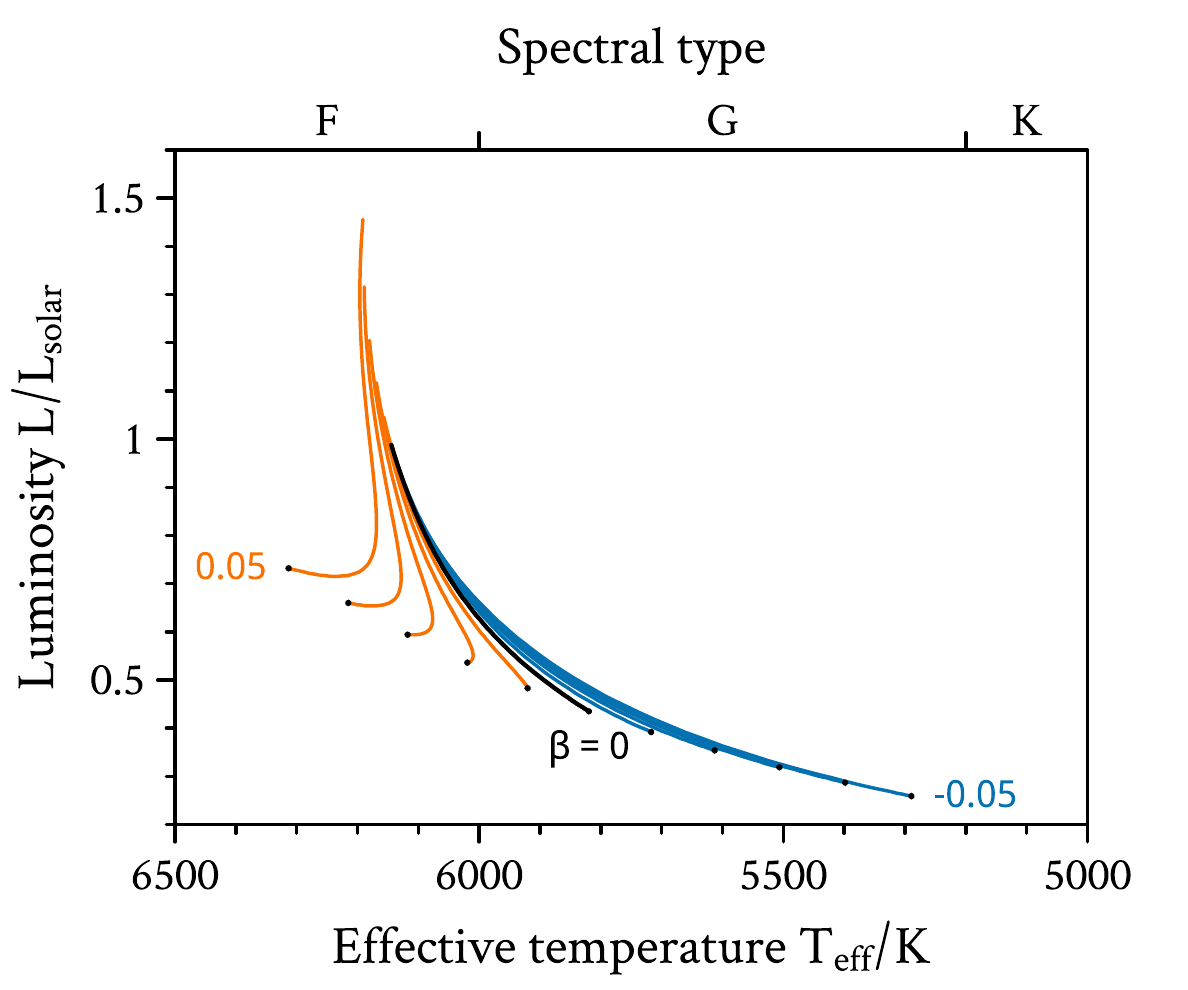}
    \caption{Theoretical evolution of a star with ${M/\text{M}_\odot=0.75}$ and ${Z=0.001}$ through the Hertzsprung--Russell diagram for varying amounts of ${\dot G}$ from the zero-age main sequence (black dots) until an age of 11~Gyr. 
    Tracks computed with positive (negative) values of $\beta$ (\emph{cf}. Equation~\ref{eq:pow-law}) are shown in orange (blue), corresponding to gravity that was stronger (weaker) in the past. \\[0.5\baselineskip]
    \label{fig:HR}} 
%
    \centering
    \includegraphics[width=\linewidth, trim={0 0 0 1.5cm}, clip]{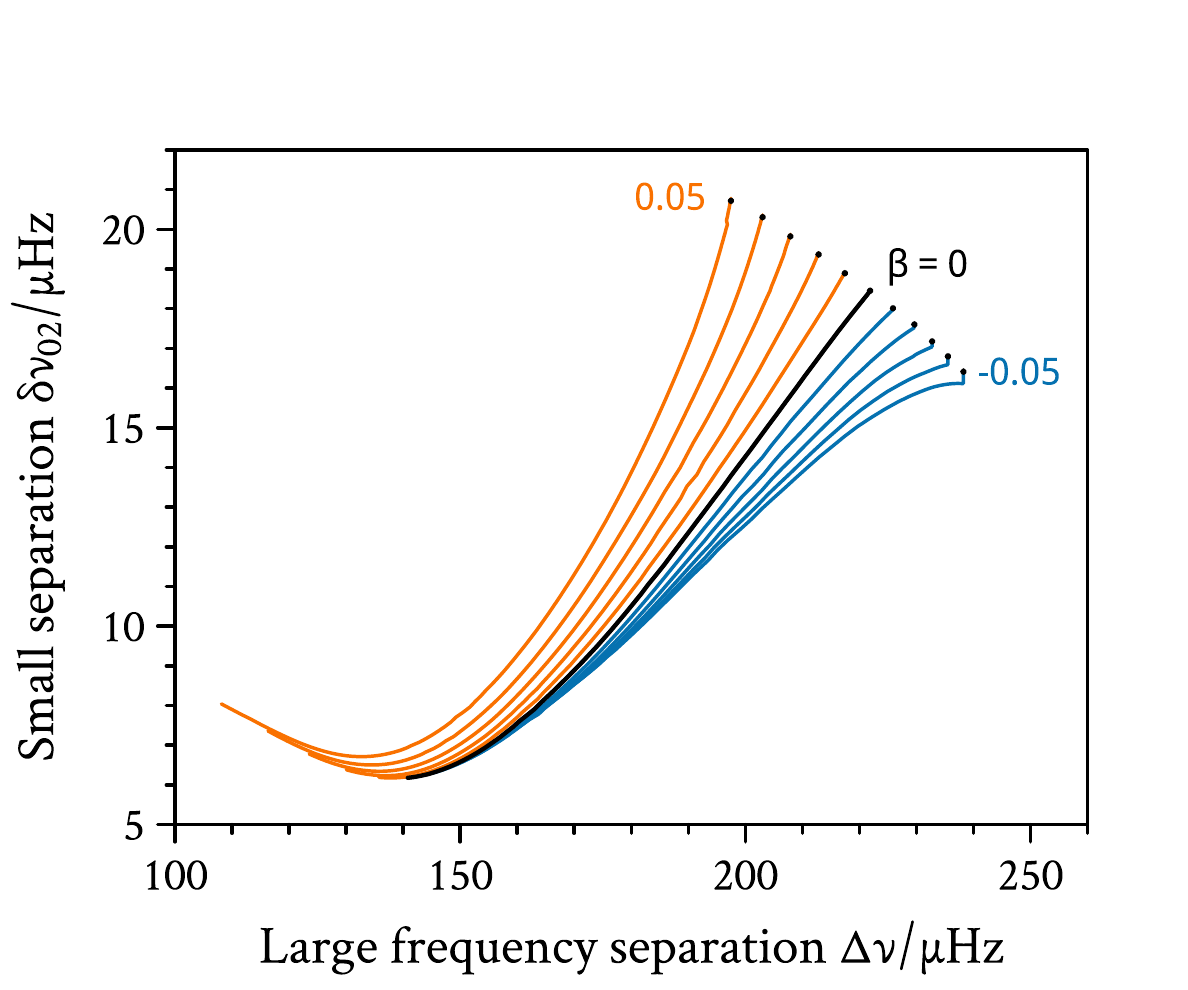}
    \caption{The same as the previous figure, now showing the theoretical evolution of the star through the asteroseismic HR diagram \citep{1988IAUS..123..295C}. 
    \label{fig:CD}}
\end{figure}

A variable gravitational constant has several consequences for stars and their evolution \citep[e.g.,][]{1977A&A....56..359M}. 
\citet{PhysRev.73.801} showed that the luminosities of stars vary as $L \propto G^7 M^5$; hence, ${\dot G \neq 0}$ directly changes the rate of stellar evolution (see Figure~\ref{fig:HR}). 
Indeed, a negative $\dot G$ has been proposed as a resolution to the faint young Sun paradox\footnote{Ironically, Teller's initial motivation for deriving this relation was to show that geological evidence is incompatible with $\dot G \neq 0$.} \citep{2014IJMPD..2342018S}. 
This modification to stellar evolution then affects acoustic stellar oscillation mode frequencies and their associated separations and ratios (see Figure~\ref{fig:CD}), as these quantities are sensitive to the composition of the stellar core. 

The star we have selected was observed in short-cadence mode (i.e., every 58.89~seconds) for nearly 3 years by \emph{Kepler}. 
Its spectroscopic data and asteroseismic frequencies were measured by \citet{2017ApJ...835..172L}, who identified 46 unique solar-like $p$-modes with spherical degrees ${\ell \leq 2}$. 
The extraordinary precision with which these measurements have been made are worthy of note: several of the modes have uncertainties smaller than 0.1~$\mu$Hz, corresponding to a relative uncertainty of approximately 0.001\%. 
The global observable parameters of this star were measured to be: 
\begin{alignat}{2}
    \text{[Fe/H]}   \;=&&~ -0.54   &\pm  0.10  ~\text{dex} \notag\\
    T_\text{eff}    \;=&&~ 5309    &\pm  77    ~\text{K} \notag\\
    \nu_{\max}      \;=&&~ 4197    &\pm 20   ~\mu\text{Hz} \\
    \Delta\nu       \;=&&~ 173.541 &\pm  0.064 ~\mu\text{Hz} \notag\\
    \delta\nu_{02}  \;=&&~ 7.901   &\pm  0.167 ~\mu\text{Hz}. \notag
\end{alignat}
The first two of these are the stellar metallicity and effective temperature. 
The quantity $\nu_{\max}$ refers to the frequency at maximum oscillation power, which is related to the surface gravity of the star \citep[e.g.,][]{2010aste.book.....A, 10.2307/j.ctt1vwmgmn}. 
The average spacing between radial oscillation mode frequencies, i.e., the large frequency separation, is given by $\Delta\nu$, and is related to the stellar mean density. 
Finally, the small frequency separation---a proxy for the main-sequence age of the star---is denoted $\delta\nu_{02}$. 
From these measurements it is clear that this star is an old, low-mass star on the main sequence. 
This description has been confirmed through detailed numerical simulations of this star by several groups \citep{2017ApJ...835..173S, 2017A&A...601A..67C, Bellinger2019a}.

\section{Methods} 
We aim to model KIC~7970740 with a time-varying gravitational constant, and determine the variations in $G$ which are empirically consistent with the stringent observational constraints that have been obtained for this star. 
As is commonly done \citep[e.g.,][]{1994ApJ...437..870D, 1998ApJ...498..871G, 1996A&A...312..345D}, we assume the gravitational constant $G$ varies over cosmic time $t$ according to a power law:
\begin{equation} \label{eq:pow-law}
    G(t) = G_0 \left(
        \frac{t_0}{t}
    \right)^\beta
\end{equation}
where ${G_0 = (6.67408 \pm 0.00031)\times 10^{-8}}$ g$^{-1}$~cm$^{3}$~s$^{-2}$ is the presently observed gravitational constant \citep{2016RvMP...88c5009M} and ${t_0=(13.799 \pm 0.021)\times 10^9}$~yr is the current age of the Universe \citep{2016A&A...594A..13P}. 
Here we seek to estimate the gravitational evolution parameter $\beta$, where a value of zero corresponds to a constant $G$. 

We use the \emph{Aarhus STellar Evolution Code} \citep[ASTEC,][]{2008Ap&SS.316...13C} to simulate the evolution of the star. 
We use the \emph{Aarhus adiabatic oscillation package} \citep[ADIPLS,][]{2008Ap&SS.316..113C} to calculate the adiabatic oscillation mode frequencies for each of the computed models. 
Example evolutionary tracks were shown in Figures~\ref{fig:HR} and \ref{fig:CD}.

In order to determine which theoretical models are consistent with the observations, we use Markov chain Monte Carlo \citep[MCMC, e.g.,][]{2010CAMCS...5...65G} 
to obtain 100\,000 samples from the posterior distribution: 
\begin{equation}
    \underbrace{p(\boldsymbol\theta | \boldsymbol D)}_{\text{posterior}}
    \propto 
    \underbrace{\mathcal{L}(\boldsymbol\theta | \boldsymbol D)}_{\text{likelihood}} 
    \cdot 
    \underbrace{p(\boldsymbol \theta)}_{\text{prior}}.
\end{equation}
Here the values ${\boldsymbol\theta = \{ \beta}, \tau, M, Y_0, Z_0, \alpha_{\text{MLT}}, t_0,  \boldsymbol S(0) \}$ are the theoretical model parameters, where $\tau$ refers to the age of the star, $M$ its mass, $Y_0$ the initial fractional abundance of helium, $Z_0$ the initial fraction of heavy mass elements, $\alpha_{\text{MLT}}$ the mixing length parameter, and $\boldsymbol S(0)$ are astrophysical S-factors of nuclear reaction rates. 
We use uniform priors on the first six of these parameters as tabulated in Table~\ref{tab:priors}. 
\mbb{These priors were adopted because previous estimates for the parameters of this star came from the analysis of the same \emph{Kepler} data, and thus normal priors would yield falsely overconfident results.} 
We adopt a normal prior on the age of the Universe as given above as well as on the astrophysical S-factors as given in Table~\ref{tab:Sfactors}. 
\mb{The posterior distribution of $\dot G/G$, as reflected primarily in the distribution of $\beta$, is the main interest of the present work.} 

\begin{table}
    \centering
    \caption{Bounds on the Uniform Prior Distributions for the Input Parameters to the Stellar Evolution Simulations \label{tab:priors}}
    \begin{tabular}{c|ccc}\hline
        Parameter & Minimum & Maximum & Unit \\\hline\hline
        $\beta$ & -0.2 & 0.2 & -- \\
        $\tau$ & 0 & $t_0$ & Gyr \\
        $M$ & 0.5 & 1 & M$_\odot$ \\ 
        $Y_0$ & 0.2 & 0.4 & -- \\
        $Z_0$ & 0.001 & 0.02 & -- \\ 
        $\alpha_{\text{MLT}}$ & 0.2 & 2.5 & -- \\\hline
    \end{tabular}
\end{table}

\begin{table}
    \centering
    \caption{Astrophysical S-factors of Nuclear Reactions \label{tab:Sfactors}}
    \begin{tabular}{l p{7mm}p{0.4mm}l}\hline
        Reaction & \multicolumn{3}{c}{$S(0)$/[keV b]} \\\hline\hline
        \hphantom{$^{14}$}$p$($p$, e+ $\nu$)$d$ & $4.01 \, (1$&$\pm$&$0.01) \times 10^{-22}$ \\
        \hphantom{$^{1}$}$^3$He($^3$He, 2$p$)$^4$He & $5.21 \, (1$&$\pm$&$0.05) \times 10^3$ \\
        \hphantom{$^{1}$}$^3$He($^4$He, $\gamma$)$^7$Be & $0.56 \, (1$&$\pm$&$0.05)$ \\
        \hphantom{$^{1}$}$^7$Be($p$, $\gamma$)$^8$B & $2.08 \, (1$&$\pm$&$0.08) \times 10^{-2}$ \\
        $^{14}$N($p$, $\gamma$)$^{15}$O & $1.66 \, (1$&$\pm$&$0.07)$ \\
        \hline
    \end{tabular}\\[0.5\baselineskip]
    \textbf{Note.} All values obtained from \citealt{2011RvMP...83..195A}. 
\end{table}

The values ${\boldsymbol D = \{T_{\text{eff}}}, \text{[Fe/H]}, \mathbf r_{102} \}$ are the observational data, the lattermost of which is a length-25 sequence comprised of $\mathbf r_{10}$ and $\mathbf r_{02}$ asteroseismic frequency separation ratios \citep{2003A&A...411..215R}. 
These are defined as: 
\begin{align}\notag
    r_{10}(n) &= \dfrac%
        {\nu_{n-1, 1} - 4\nu_{n,0} + 6\nu_{n,1} - 4\nu_{n+1,0} + \nu_{n+1,1}}%
        {\nu_{n,0} - 8\nu_{n+1,0}}\\
    r_{02}(n) &= \dfrac{\nu_{n, 0} - \nu_{n-1, 2}}{\nu_{n,1} - \nu_{n-1, 1}}
\end{align}
where $\nu_{n,\ell}$ refers to the frequency of the mode with radial order $n$ and spherical degree $\ell$. 
These quantities are useful because they probe the interior structure of the star and are insensitive to the near-surface layers. 

The likelihood of the observed data for a given set of input parameters is given by
\begin{equation}
    \mathcal{L}(\boldsymbol\theta | \boldsymbol D) 
    \propto 
    \exp\left(-\frac{\chi^2}{2}\right)
\end{equation}
where the goodness-of-fit $\chi^2$ in this case is
\begin{align}
    \chi^2
    = 
    \mathbf{R}^{\text{T}} \boldsymbol{\Sigma}^{-1} \mathbf{R}%
    ,\qquad 
    R_i 
    = 
    D_i - A_i(\boldsymbol \theta). \label{eq:chi2}
\end{align}
Here $\mathbf \Sigma$ 
is the full variance--covariance matrix for the observations, which accounts for the fact the observed asteroseismic frequency ratios are correlated \citep{2018arXiv180807556R}; and $\mathbf A$ is the result of calling ASTEC and ADIPLS with the given model parameters $\boldsymbol \theta$. 
\mbb{It is worthy of mention that previous MCMC asteroseismic modeling has considered at most a four-dimensional parameter space \citep[see, e.g.,][]{2012MNRAS.427.1847B, 2018ASSP...49..149L, 2019MNRAS.484..771R}. 
With twelve dimensions, this is, to our knowledge, the most complex asteroseismic modeling performed to date.}

\section{Results \& Conclusions} 

\begin{figure}
    \centering
    \includegraphics[width=\linewidth, trim={0 0 0 0.5cm}, clip]{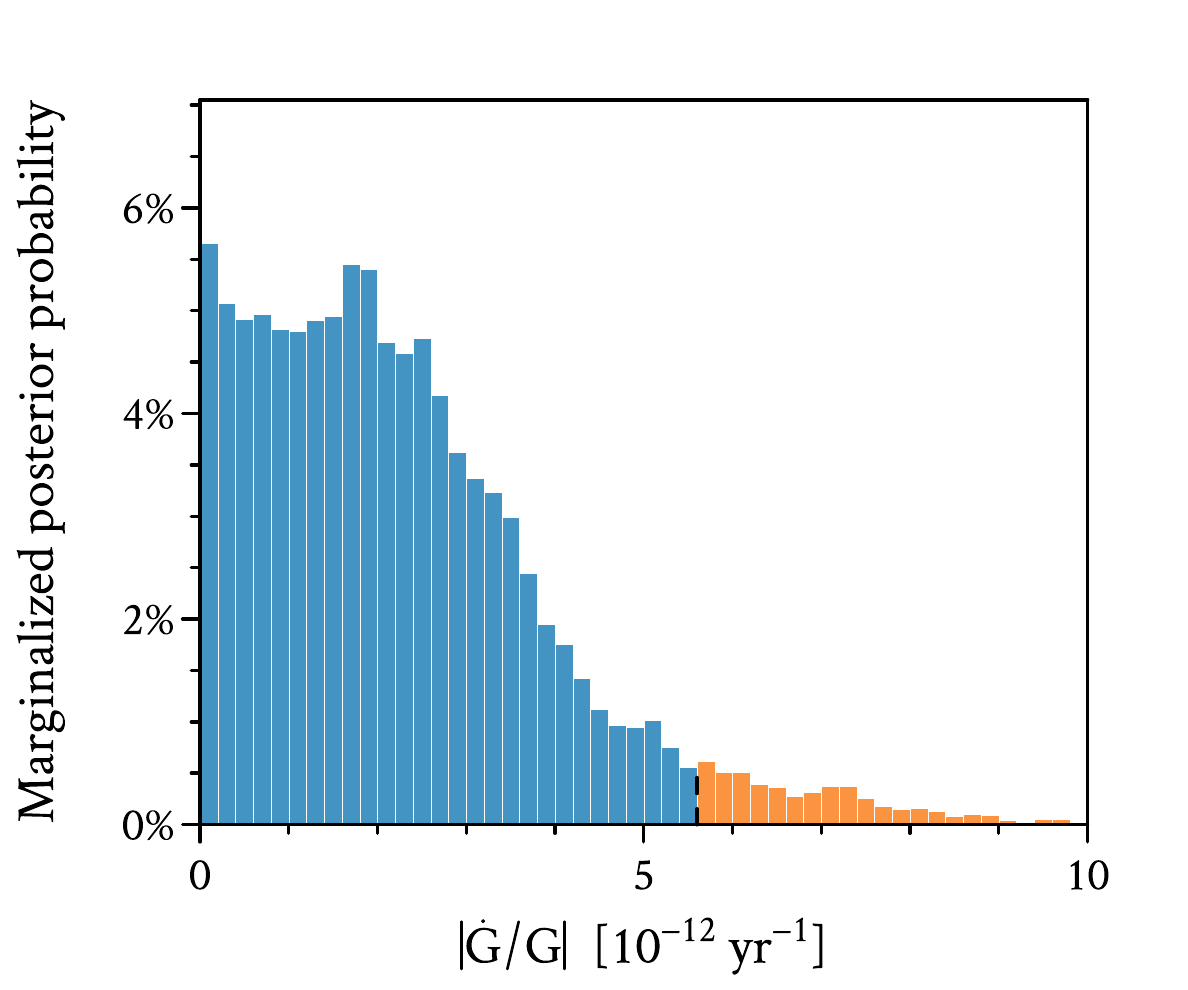}
    \caption{\mb{Histogram of MCMC samples showing} the upper bound on the possible variation of $|\dot G/G|$. The 95\% credible interval is in blue. Values closer to zero are in support of general relativity; values farther from zero are in support of modified gravity. 
    \label{fig:upper-bound}} 
\end{figure}

The procedure outlined in the previous section yields several results. 
The main result is the value of the gravitational evolution parameter, which we find to be ${\beta = \mbbb{0.017 \pm 0.035}}$. 
We also infer from this analysis an estimate of the age of the Universe, which we find to be ${t_0 = \mbbb{13.797 \pm 0.019}}$~Gyr. 
Combining these two quantities yields a rate of change in $G$ of 
\begin{equation}
    \dot{G}/G = \beta/t_0 = \mbbb{(1.2 \pm 2.6)} \times 10^{-12}~\text{yr}^{-1}.
\end{equation}
Hence we find no evidence for a variable gravitational constant. 
We furthermore place a 95\% upper bound on the absolute variation 
\begin{equation}
    |\dot{G}/G| \leq \mbbb{5.6} \times 10^{-12}\;\text{yr}^{-1}
\end{equation}
as visualized in Figure~\ref{fig:upper-bound}. 
The posterior estimates for the five nuclear reaction rates are consistent with their prior values. 
We tested this procedure under two assumptions of the solar composition: the \citet[][``GS98'']{1998SSRv...85..161G} values and the \citet[][``AGSS09'']{2009ARA&A..47..481A} values, and found the results to be the same. 
\mb{These results are stronger than those from big bang nucleosynthesis, but probe less time; and weaker than those from helioseismology, but probe more than twice as much time.} 

Lastly, we obtain new estimates for the stellar parameters of KIC~7970740: 
\begin{alignat}{2}
    \tau   \;=&&~ \mbbb{10.9}   &\pm  \mbbb{1.2}  ~\text{Gyr} \notag\\
    M    \;=&&~ \mbbb{0.725}    &\pm  \mbbb{0.043}    ~\text{M}_\odot \notag\\
    Y_0      \;=&&~ \mbbb{0.252}    &\pm \mbbb{0.035} \\
    Z_0       \;=&&~ \mbbb{0.0058} &\pm  \mbbb{0.0012} \notag\\
    \alpha_{\text{MLT}}  \;=&&~ \mbbb{1.89}   &\pm  \mbbb{0.23}. \notag
\end{alignat}
These values are in good agreement with those presented by \citet{2017ApJ...835..173S}, who found for this star a mass of ${0.728 \pm 0.020~\text{M}_\odot}$ and an age of $12.9 \pm 1.6$~Gyr. 
It is worthy of note that the mean posterior value of the initial helium abundance of this star is above the primordial helium abundance $Y_p = 0.2463$ inferred by the \emph{Planck} mission \citep{2014JCAP...10..050C}.

\mbb{Investigation into the constancy of G is still a very active area of inquiry spanning a wide range of domains in astrophysics. 
This work lays a bridge between asteroseismology and these other disciplines by enabling the use of individual stars for obtaining constraints at every age.}  
In the future, it will be interesting to apply this technique to an ensemble of stars, which should yield an even stronger result. 
In addition, it will be interesting to use asteroseismology to constrain the variation of other values that are thought to be constant, such as the fine structure constant \citep{2006ESASP.617E..37B}.

\acknowledgements{} 
    We thank Margarida Cunha, Martin Bo Nielsen, and the anonymous referee for useful comments and suggestions. 
    Funding for the Stellar Astrophysics Centre is provided by The Danish National Research Foundation (Grant agreement no.: DNRF106). 
    The numerical results presented in this work were obtained at the Centre for Scientific Computing, Aarhus. 
\software{} Python, emcee \citep{emcee}, R \citep{r}, ASTEC, and ADIPLS \citep{2008Ap&SS.316...13C, 2008Ap&SS.316..113C}

\bibliographystyle{aasjournal.bst}
\bibliography{main}

\begin{thebibliography}{}
\expandafter\ifx\csname natexlab\endcsname\relax\def\natexlab#1{#1}\fi
\providecommand{\url}[1]{\href{#1}{#1}}
\providecommand{\dodoi}[1]{doi:~\href{http://doi.org/#1}{\nolinkurl{#1}}}
\providecommand{\doeprint}[1]{\href{http://ascl.net/#1}{\nolinkurl{http://ascl.net/#1}}}
\providecommand{\doarXiv}[1]{\href{https://arxiv.org/abs/#1}{\nolinkurl{https://arxiv.org/abs/#1}}}

\bibitem[{{Accetta} {et~al.}(1990){Accetta}, {Krauss}, \&
  {Romanelli}}]{1990PhLB..248..146A}
{Accetta}, F.~S., {Krauss}, L.~M., \& {Romanelli}, P. 1990, Physics Letters B,
  248, 146, \dodoi{10.1016/0370-2693(90)90029-6}

\bibitem[{{Adelberger} {et~al.}(2011){Adelberger}, {Garc{\'{\i}}a},
  {Robertson}, {Snover}, {Balantekin}, {Heeger}, {Ramsey-Musolf}, {Bemmerer},
  {Junghans}, {Bertulani}, {Chen}, {Costantini}, {Prati}, {Couder},
  {Uberseder}, {Wiescher}, {Cyburt}, {Davids}, {Freedman}, {Gai}, {Gazit},
  {Gialanella}, {Imbriani}, {Greife}, {Hass}, {Haxton}, {Itahashi}, {Kubodera},
  {Langanke}, {Leitner}, {Leitner}, {Vetter}, {Winslow}, {Marcucci},
  {Motobayashi}, {Mukhamedzhanov}, {Tribble}, {Nollett}, {Nunes}, {Park},
  {Parker}, {Schiavilla}, {Simpson}, {Spitaleri}, {Strieder}, {Trautvetter},
  {Suemmerer}, \& {Typel}}]{2011RvMP...83..195A}
{Adelberger}, E.~G., {Garc{\'{\i}}a}, A., {Robertson}, R.~G.~H., {et~al.} 2011,
  RMP, 83, 195, \dodoi{10.1103/RevModPhys.83.195}

\bibitem[{{Aerts} {et~al.}(2010){Aerts}, {Christensen-Dalsgaard}, \&
  {Kurtz}}]{2010aste.book.....A}
{Aerts}, C., {Christensen-Dalsgaard}, J., \& {Kurtz}, D.~W. 2010,
  {Asteroseismology} (Springer Science)

\bibitem[{{Asplund} {et~al.}(2009){Asplund}, {Grevesse}, {Sauval}, \&
  {Scott}}]{2009ARA&A..47..481A}
{Asplund}, M., {Grevesse}, N., {Sauval}, A.~J., \& {Scott}, P. 2009, \araa, 47,
  481, \dodoi{10.1146/annurev.astro.46.060407.145222}

\bibitem[{Basu \& Chaplin(2017)}]{10.2307/j.ctt1vwmgmn}
Basu, S., \& Chaplin, W.~J. 2017, Asteroseismic Data Analysis (Princeton
  University Press).
\newblock \url{http://www.jstor.org/stable/j.ctt1vwmgmn}

\bibitem[{{Bazot} {et~al.}(2012){Bazot}, {Bourguignon}, \&
  {Christensen-Dalsgaard}}]{2012MNRAS.427.1847B}
{Bazot}, M., {Bourguignon}, S., \& {Christensen-Dalsgaard}, J. 2012, \mnras,
  427, 1847, \dodoi{10.1111/j.1365-2966.2012.21818.x}

\bibitem[{{Bellinger} {et~al.}(2019){Bellinger}, {Hekker, S.}, {Angelou, G.
  C.}, {Stokholm, A.}, \& {Basu, S.}}]{Bellinger2019a}
{Bellinger}, E.~P., {Hekker, S.}, {Angelou, G. C.}, {Stokholm, A.}, \& {Basu,
  S.} 2019, \aap, 622, arXiv:1812.06979, \dodoi{10.1051/0004-6361/201834461}

\bibitem[{{Bonanno} \& {Schlattl}(2006)}]{2006ESASP.617E..37B}
{Bonanno}, A., \& {Schlattl}, H. 2006, in ESA Special Publication, Vol. 617,
  SOHO-17, 37

\bibitem[{{Borucki} {et~al.}(2010){Borucki}, {Koch}, {Basri}, {Batalha},
  {Brown}, {Caldwell}, {Caldwell}, {Christensen-Dalsgaard}, {Cochran},
  {DeVore}, {Dunham}, {Dupree}, {Gautier}, {Geary}, {Gilliland}, {Gould},
  {Howell}, {Jenkins}, {Kondo}, {Latham}, {Marcy}, {Meibom}, {Kjeldsen},
  {Lissauer}, {Monet}, {Morrison}, {Sasselov}, {Tarter}, {Boss}, {Brownlee},
  {Owen}, {Buzasi}, {Charbonneau}, {Doyle}, {Fortney}, {Ford}, {Holman},
  {Seager}, {Steffen}, {Welsh}, {Rowe}, {Anderson}, {Buchhave}, {Ciardi},
  {Walkowicz}, {Sherry}, {Horch}, {Isaacson}, {Everett}, {Fischer}, {Torres},
  {Johnson}, {Endl}, {MacQueen}, {Bryson}, {Dotson}, {Haas}, {Kolodziejczak},
  {Van Cleve}, {Chandrasekaran}, {Twicken}, {Quintana}, {Clarke}, {Allen},
  {Li}, {Wu}, {Tenenbaum}, {Verner}, {Bruhweiler}, {Barnes}, \&
  {Prsa}}]{2010Sci...327..977B}
{Borucki}, W.~J., {Koch}, D., {Basri}, G., {et~al.} 2010, Science, 327, 977,
  \dodoi{10.1126/science.1185402}

\bibitem[{{Chiba}(2011)}]{2011PThPh.126..993C}
{Chiba}, T. 2011, Progress of Theoretical Physics, 126, 993,
  \dodoi{10.1143/PTP.126.993}

\bibitem[{{Christensen-Dalsgaard}(1988)}]{1988IAUS..123..295C}
{Christensen-Dalsgaard}, J. 1988, in IAU Symposium, Vol. 123, Advances in
  Helio- and Asteroseismology, ed. J.~{Christensen-Dalsgaard} \& S.~{Frandsen},
  295

\bibitem[{{Christensen-Dalsgaard}(2008{\natexlab{a}})}]{2008Ap&SS.316...13C}
{Christensen-Dalsgaard}, J. 2008{\natexlab{a}}, \apss, 316, 13,
  \dodoi{10.1007/s10509-007-9675-5}

\bibitem[{{Christensen-Dalsgaard}(2008{\natexlab{b}})}]{2008Ap&SS.316..113C}
---. 2008{\natexlab{b}}, \apss, 316, 113, \dodoi{10.1007/s10509-007-9689-z}

\bibitem[{{Coc} {et~al.}(2014){Coc}, {Uzan}, \&
  {Vangioni}}]{2014JCAP...10..050C}
{Coc}, A., {Uzan}, J.-P., \& {Vangioni}, E. 2014, \jcap, 2014, 050,
  \dodoi{10.1088/1475-7516/2014/10/050}

\bibitem[{{C{\'o}rsico} {et~al.}(2013){C{\'o}rsico}, {Althaus},
  {Garc{\'{\i}}a-Berro}, \& {Romero}}]{2013JCAP...06..032C}
{C{\'o}rsico}, A.~H., {Althaus}, L.~G., {Garc{\'{\i}}a-Berro}, E., \& {Romero},
  A.~D. 2013, \jcap, 6, 032, \dodoi{10.1088/1475-7516/2013/06/032}

\bibitem[{{Creevey} {et~al.}(2017){Creevey}, {Metcalfe}, {Schultheis},
  {Salabert}, {Bazot}, {Th{\'e}venin}, {Mathur}, {Xu}, \&
  {Garc{\'{\i}}a}}]{2017A&A...601A..67C}
{Creevey}, O.~L., {Metcalfe}, T.~S., {Schultheis}, M., {et~al.} 2017, \aap,
  601, A67, \dodoi{10.1051/0004-6361/201629496}

\bibitem[{{Damour} \& {Taylor}(1991)}]{1991ApJ...366..501D}
{Damour}, T., \& {Taylor}, J.~H. 1991, \apj, 366, 501, \dodoi{10.1086/169585}

\bibitem[{{degl'Innocenti} {et~al.}(1996){degl'Innocenti}, {Fiorentini},
  {Raffelt}, {Ricci}, \& {Weiss}}]{1996A&A...312..345D}
{degl'Innocenti}, S., {Fiorentini}, G., {Raffelt}, G.~G., {Ricci}, B., \&
  {Weiss}, A. 1996, \aap, 312, 345

\bibitem[{{Demarque} {et~al.}(1994){Demarque}, {Krauss}, {Guenther}, \&
  {Nydam}}]{1994ApJ...437..870D}
{Demarque}, P., {Krauss}, L.~M., {Guenther}, D.~B., \& {Nydam}, D. 1994, \apj,
  437, 870, \dodoi{10.1086/175048}

\bibitem[{{Dirac}(1937)}]{1937Natur.139..323D}
{Dirac}, P.~A.~M. 1937, \nat, 139, 323, \dodoi{10.1038/139323a0}

\bibitem[{{Foreman-Mackey} {et~al.}(2013){Foreman-Mackey}, {Hogg}, {Lang}, \&
  {Goodman}}]{emcee}
{Foreman-Mackey}, D., {Hogg}, D.~W., {Lang}, D., \& {Goodman}, J. 2013, PASP,
  125, 306, \dodoi{10.1086/670067}

\bibitem[{{Garc{\'{\i}}a-Berro} {et~al.}(2011){Garc{\'{\i}}a-Berro},
  {Lor{\'e}n-Aguilar}, {Torres}, {Althaus}, \& {Isern}}]{2011JCAP...05..021G}
{Garc{\'{\i}}a-Berro}, E., {Lor{\'e}n-Aguilar}, P., {Torres}, S., {Althaus},
  L.~G., \& {Isern}, J. 2011, \jcap, 5, 021,
  \dodoi{10.1088/1475-7516/2011/05/021}

\bibitem[{{Genova} {et~al.}(2018){Genova}, {Mazarico}, {Goossens}, {Lemoine},
  {Neumann}, {Smith}, \& {Zuber}}]{2018NatCo...9..289G}
{Genova}, A., {Mazarico}, E., {Goossens}, S., {et~al.} 2018, Nature Comm., 9,
  289, \dodoi{10.1038/s41467-017-02558-1}

\bibitem[{{Goodman} \& {Weare}(2010)}]{2010CAMCS...5...65G}
{Goodman}, J., \& {Weare}, J. 2010, Comm. in Appl. Math. and Comp. Sci., 5, 65,
  \dodoi{10.2140/camcos.2010.5.65}

\bibitem[{{Grevesse} \& {Sauval}(1998)}]{1998SSRv...85..161G}
{Grevesse}, N., \& {Sauval}, A.~J. 1998, \ssr, 85, 161,
  \dodoi{10.1023/A:1005161325181}

\bibitem[{{Guenther} {et~al.}(1998){Guenther}, {Krauss}, \&
  {Demarque}}]{1998ApJ...498..871G}
{Guenther}, D.~B., {Krauss}, L.~M., \& {Demarque}, P. 1998, \apj, 498, 871,
  \dodoi{10.1086/305567}

\bibitem[{Hellings {et~al.}(1983)Hellings, Adams, Anderson, Keesey, Lau,
  Standish, Canuto, \& Goldman}]{PhysRevLett.51.1609}
Hellings, R.~W., Adams, P.~J., Anderson, J.~D., {et~al.} 1983, Phys. Rev.
  Lett., 51, 1609, \dodoi{10.1103/PhysRevLett.51.1609}

\bibitem[{Hofmann \& M\"uller(2018)}]{0264-9381-35-3-035015}
Hofmann, F., \& M\"uller, J. 2018, Classical and Quantum Gravity, 35, 035015.
\newblock \url{http://stacks.iop.org/0264-9381/35/i=3/a=035015}

\bibitem[{{Lund} \& {Reese}(2018)}]{2018ASSP...49..149L}
{Lund}, M.~N., \& {Reese}, D.~R. 2018, Asteroseismology and Exoplanets:
  Listening to the Stars and Searching for New Worlds, 49, 149,
  \dodoi{10.1007/978-3-319-59315-9_8}

\bibitem[{{Lund} {et~al.}(2017){Lund}, {Silva Aguirre}, {Davies}, {Chaplin},
  {Christensen-Dalsgaard}, {Houdek}, {White}, {Bedding}, {Ball}, {Huber},
  {Antia}, {Lebreton}, {Latham}, {Handberg}, {Verma}, {Basu}, {Casagrande},
  {Justesen}, {Kjeldsen}, \& {Mosumgaard}}]{2017ApJ...835..172L}
{Lund}, M.~N., {Silva Aguirre}, V., {Davies}, G.~R., {et~al.} 2017, \apj, 835,
  172, \dodoi{10.3847/1538-4357/835/2/172}

\bibitem[{{Maeder}(1977)}]{1977A&A....56..359M}
{Maeder}, A. 1977, \aap, 56, 359

\bibitem[{{Masuda} \& {Suto}(2016)}]{2016PASJ...68L...5M}
{Masuda}, K., \& {Suto}, Y. 2016, \pasj, 68, L5, \dodoi{10.1093/pasj/psw017}

\bibitem[{{Mohr} {et~al.}(2016){Mohr}, {Newell}, \&
  {Taylor}}]{2016RvMP...88c5009M}
{Mohr}, P.~J., {Newell}, D.~B., \& {Taylor}, B.~N. 2016, RMP, 88, 035009,
  \dodoi{10.1103/RevModPhys.88.035009}

\bibitem[{{Murphy}(2013)}]{2013RPPh...76g6901M}
{Murphy}, T.~W. 2013, Reports on Progress in Physics, 76, 076901,
  \dodoi{10.1088/0034-4885/76/7/076901}

\bibitem[{{Nagata} {et~al.}(2004){Nagata}, {Chiba}, \&
  {Sugiyama}}]{2004PhRvD..69h3512N}
{Nagata}, R., {Chiba}, T., \& {Sugiyama}, N. 2004, \prd, 69, 083512,
  \dodoi{10.1103/PhysRevD.69.083512}

\bibitem[{{Planck Collaboration} {et~al.}(2016){Planck Collaboration}, {Ade},
  {Aghanim}, {Arnaud}, {Ashdown}, {Aumont}, {Baccigalupi}, {Banday},
  {Barreiro}, {Bartlett}, \& et~al.}]{2016A&A...594A..13P}
{Planck Collaboration}, {Ade}, P.~A.~R., {Aghanim}, N., {et~al.} 2016, \aap,
  594, A13, \dodoi{10.1051/0004-6361/201525830}

\bibitem[{{R Core Team}(2014)}]{r}
{R Core Team}. 2014, R, R Foundation.
\newblock \url{http://www.R-project.org/}

\bibitem[{{Rendle} {et~al.}(2019){Rendle}, {Buldgen}, {Miglio}, {Reese},
  {Noels}, {Davies}, {Campante}, {Chaplin}, {Lund}, {Kuszlewicz}, {Scott},
  {Scuflaire}, {Ball}, {Smetana}, \& {Nsamba}}]{2019MNRAS.484..771R}
{Rendle}, B.~M., {Buldgen}, G., {Miglio}, A., {et~al.} 2019, \mnras, 484, 771,
  \dodoi{10.1093/mnras/stz031}

\bibitem[{{Roxburgh}(2018)}]{2018arXiv180807556R}
{Roxburgh}, I.~W. 2018, arXiv e-prints.
\newblock \doarXiv{1808.07556}

\bibitem[{{Roxburgh} \& {Vorontsov}(2003)}]{2003A&A...411..215R}
{Roxburgh}, I.~W., \& {Vorontsov}, S.~V. 2003, \aap, 411, 215,
  \dodoi{10.1051/0004-6361:20031318}

\bibitem[{{Sahni} \& {Shtanov}(2014)}]{2014IJMPD..2342018S}
{Sahni}, V., \& {Shtanov}, Y. 2014, Int. J. Mod. Phys. D, 23, 1442018,
  \dodoi{10.1142/S0218271814420188}

\bibitem[{{Silva Aguirre} {et~al.}(2017){Silva Aguirre}, {Lund}, {Antia},
  {Ball}, {Basu}, {Christensen-Dalsgaard}, {Lebreton}, {Reese}, {Verma},
  {Casagrande}, {Justesen}, {Mosumgaard}, {Chaplin}, {Bedding}, {Davies},
  {Handberg}, {Houdek}, {Huber}, {Kjeldsen}, {Latham}, {White}, {Coelho},
  {Miglio}, \& {Rendle}}]{2017ApJ...835..173S}
{Silva Aguirre}, V., {Lund}, M.~N., {Antia}, H.~M., {et~al.} 2017, \apj, 835,
  173, \dodoi{10.3847/1538-4357/835/2/173}

\bibitem[{{Smullin} \& {Fiocco}(1962)}]{1962Natur.194.1267S}
{Smullin}, L.~D., \& {Fiocco}, G. 1962, \nat, 194, 1267,
  \dodoi{10.1038/1941267a0}

\bibitem[{Teller(1948)}]{PhysRev.73.801}
Teller, E. 1948, Phys. Rev., 73, 801, \dodoi{10.1103/PhysRev.73.801}

\bibitem[{{Uzan}(2003)}]{2003RvMP...75..403U}
{Uzan}, J.-P. 2003, RMP, 75, 403, \dodoi{10.1103/RevModPhys.75.403}

\bibitem[{Uzan(2011)}]{Uzan2011}
Uzan, J.-P. 2011, Living Reviews in Relativity, 14, 2,
  \dodoi{10.12942/lrr-2011-2}

\bibitem[{{Zhu} {et~al.}(2019){Zhu}, {Desvignes}, {Wex}, {Caballero},
  {Champion}, {Demorest}, {Ellis}, {Janssen}, {Kramer}, {Krieger}, {Lentati},
  {Nice}, {Ransom}, {Stairs}, {Stappers}, {Verbiest}, {Arzoumanian}, {Bassa},
  {Burgay}, {Cognard}, {Crowter}, {Dolch}, {Ferdman}, {Fonseca}, {Gonzalez},
  {Graikou}, {Guillemot}, {Hessels}, {Jessner}, {Jones}, {Jones}, {Jordan},
  {Karuppusamy}, {Lam}, {Lazaridis}, {Lazarus}, {Lee}, {Levin}, {Liu}, {Lyne},
  {McKee}, {McLaughlin}, {Os{\l}owski}, {Pennucci}, {Perrodin}, {Possenti},
  {Sanidas}, {Shaifullah}, {Smits}, {Stovall}, {Swiggum}, {Theureau}, \&
  {Tiburzi}}]{2019MNRAS.482.3249Z}
{Zhu}, W.~W., {Desvignes}, G., {Wex}, N., {et~al.} 2019, \mnras, 482, 3249,
  \dodoi{10.1093/mnras/sty2905}

\end{thebibliography}

\end{document}